\documentstyle[11pt]{article}
\catcode`\@=11
\begin{document} \openup6pt

\title{CHAOTIC INFLATIONARY UNIVERSE ON  BRANE}

\author{B. C. Paul\thanks{Electronic mail : bcpaul@iucaa.ernet.in} \\
	Physics Department, North Bengal University, \\
Siliguri, DIST. : Darjeeling, Pin : 734 430, West Bengal, INDIA \\
and \\
Inter-University Centre for Astronomy and Astrophysics \\
P.O. Box : 4, Ganeshkhind, Pune - 411 007, India}

\date{}

\maketitle

\vspace{0.5in}

\begin{abstract}

The chaotic inflationary model of the early universe, proposed by Linde is explored in  the brane world   considering  matter described by a minimally coupled self interacting scalar field. We obtain cosmological solutions which admit evolution of a universe  either from a singularity or without a singularity. It is found that a very weakly coupled self-interacting scalar field is necessary for a quartic type potential in the brane world model compared to that necessary in general relativity. In the brane world sufficient inflation may be obtained even with an initial  scalar field having value less than the Planck scale. It is found 
that if the universe is kinetic energy dominated to begin with, it transits to an inflationary stage subsequently.
\end{abstract}

\vspace{0.2cm}

PACS number(s) : 04.50.+h, 98.80.Cq

\vspace{4.5cm}

\pagebreak

{\bf I. INTRODUCTION }
  
It is now generally believed that the Einstein's general theory of relativity (GTR) needs to be modified at a very high energy scale. Such a high energy perhaps was available in the universe shortly after the big bang. 
 It  is likely that the GTR may reduce to a limiting case of a more general theory. To accommodate such a theory one requires a  framework where   gravity may be quantized. But a consistent quantum theory of gravity is yet to emerge. The developmental works [1, 2] in superstring and 
 M-theory  in the last few years have been recognised as important to build a consistent theory of quantum gravity. However, those theories need space-time dimensions more than the usual four dimensions for its consistent formulation. There have been an important paradigm shift in recent years to explore a higher dimensional universe from that of the previous approaches initiated by Kaluza and Klein [3] and others [4]. In modern higher dimensional scenario, the fields of the standard model are considered to be confined to 3 + 1 dimensional hyper-surface (referred to as 3 brane) embedded in  higher dimensional spacetime but the gravitational field may propagate through the bulk dimensions perpendicular to the brane which is referred  to as the braneworld.  According to 
Randall and Sundrum [2] although the extra dimensions is not compact, four dimensional Newtonian gravity is recovered in five dimensional anti-de Sitter spacetime ($AdS_{5}$) in the low energy limit.  It is also found that the heirarchy problem in particle physics gets resolved in brane world. In the last couple of years a number of literature appeared in the braneworld cosmology  which addressed  isotropic [5] as well as anisotropic [6] brane  cosmologies, 
since gravity in braneworld may differ considerably from the conventional four dimensional gravity.  
It is now generally accepted that inflation is an essential ingredient in modern cosmology which one obtains using the theories of particle physics. During the inflationary epoch, the scale factor of the universe grew exponentially (or quasiexponentially) allowing a small causally coherent region to become big enough to be identified with the present observable universe. 
Consequently the inflationary universe scenario  provides a satisfactory solution to some of the conceptual issues in cosmology not understood in the standard Big-Bang theory.    Maartens { \it et al.} [7]  studied chaotic inflation on the brane using a massive scalar field. It was  found that the Friedmann constraint equation which led  to a faster Hubble expansion at high energies and a more strongly damped evolving scalar field, help very much to attain easily the condition for slow-roll inflation
for  a given potential.  It was also found  that the inflationary scenario in brane cosmology has the desired scale invariant Harrison-Zeldovich spectrum  for density perturbation in accordance
 with the   COBE data.    Chaotic  inflation on the FRW brane [7] was obtained with an assumption that the potential energy dominates over that of the kinetic energy of the scalar field.  In GTR, Linde [8] has shown that chaotic scenario  can be realised even when scalar field $\phi$ obeys
 $ \frac{1}{2} \dot{\phi}^{2} >> V(\phi)$ and attains the required condition for inflation in course of its evolution in FRW universe. In a subsequent paper Paul {\it et al.} [9] shown that Linde's chaotic scenario is more general and can be accommodated even in an anisotropic universe.
Chaotic model of the early universe is based on a single scalar field which can be formulated either as a theory of a massless self-interacting scalar field with quartic self-coupling ($b_{o}$) or a non-interacting massless scalar field with mass $m$. For a realistic inflationary universe scenario, one fixes either the mass or quartic coupling, $m \sim 10^{-5} M_{4}$ ($M_{4}$ represents the Planck mass) or $b_{o} \sim 10^{- 12}$ weakly  coupled field. For a sufficient inflation one requires initial values of the inflaton field $\phi_{i} > few \, M_{4}$. However, it is troublesome to assume $\phi_{i} > few \, M_{4}$ and use only the quadratic or quartic potential terms, as in these cases one is implicitly assuming that all non-renormalizable terms are absent or severly suppressed [10]. However, considering two or more scalar field in the theory may improve the situation as it releive us from the problem of extremely large initial field values. Linde [11], Kanti and Olive [12] shown that the fine tunning of the quartic coupling $b_{o} \sim 10^{- 12}$ could be relaxed in the framework of a four dimensional theory of multiple scalar fields.  They argued that the source of a large number of nearly identical scalar fields could be the result of compactification of a large extra dimension. It was also shown [13] that the multiplicity of the scalar field may resolve the problem of large initial condition necessary for chaotic inflation.
 In
this paper we consider self-interacting scalar field to obtain brane cosmological solutions and look for inflation in the early universe considering the initial conditions for inflation in one case and in another case we look for inflation relaxing the conditions for inflation.
 
The paper is organized as follows : The field equations on the brane are written down in sec. II. In sec. III we obtain cosmological solution and in 
 sec IV.  a brief discussion.

{\bf II. BRANE  FIELD EQUATION }

In the five dimensional (bulk) space-time the Einstein's field equation is
given by 
\begin{equation}
G_{AB}^{(5)} = \tilde{\kappa}^{2} \left[ - g_{AB}^{(5)} \Lambda_{(5)} + 
 T_{AB}^{(5)} \right] 
\end{equation}
with $T_{AB}^{(5)} = \delta (y) [ - \lambda g_{AB} + T_{AB} ]$. 
Here $\kappa_{(5)}$ represents the  five dimensional gravitational 
coupling constant, $g_{AB}^{(5)}$, 
$G_{AB}^{(5)} $ and  $\Lambda_{(5)} $ are the metric, Einstein tensor and 
the cosmological constant of the bulk space-time respectively, $T_{AB}$ is 
the matter energy momentum tensor. We have  $\tilde{\kappa}^{2} = 
\frac{8 \pi}{M_{5}^{3}}$.  A natural  choice of coordinates is $x^{A} = ( x^{\mu},
 y ) $ where $x^{\mu} = (t, x^{i})$ are space-time coordinates on the brane. 
The the upper case Latin letters
($A, B, ... = 0, ..., 4$)  represents  coordinate indices in 
the bulk spacetime, the Greek letters  ($\mu, \nu, ... = 0, ..., 3$)
for the coordinate indices in the four dimensional
spacetime and the
small case latin letters  ($i, j = 1, 2, 3 $) for three space. 
The space-like hypersurface $x^{4} = y = 0$ gives the brane world and
 $g_{AB}$ is its induced metric, $\lambda $ is the tension of the brane 
which is assumed to be positive in order to recover conventional general
theory of gravity (GTR) on the brane. The bulk cosmological constant 
$\Lambda_{(5)} $ is negative and represents the five dimensional cosmological
constant.

The field equations induced on the brane are derived using geometric 
approach [14] leading to new terms which carry bulk effects on the
 brane. The modified dynamical equations on the brane is
\begin{equation}
G_{\mu \nu} = - \Lambda g_{\mu \nu} +   \kappa^{2} T_{\mu \nu} 
 + \tilde{\kappa}^{4} S_{\mu \nu} - E_{\mu \nu},
\end{equation}
where the effective cosmological 
constant $ \Lambda $ and the four dimensional constant $ \kappa $ 
 on the brane are given by
\[
\Lambda = \frac{ \Lambda_{5} }{2}  \left[ 
\left( \frac{\lambda}{\lambda_{c}} \right)^{2} - 1 \right] 
\]
\begin{equation}
\kappa^{2}  = \frac{1}{6} \lambda \; \tilde{\kappa}^{4},
\end{equation}
respectively, and  $\lambda_{c}$ is the critical brane tension which is 
\begin{equation}
\lambda_{c} = 6 \frac{\Lambda_{5}}{ \tilde{\kappa}^{2} } .
\end{equation}
However one can make the effective four dimensional cosmological 
constant zero by a choice of the brane tension.  The 
extra dimensional corrections to the 
Einstein equations on the brane are of two types and are given by : 

$\bullet $
$S_{\mu \nu} $ :  quadratic in the matter variables which is
\begin{equation}
S_{\mu \nu} =  \frac{1}{12} T T_{\mu \nu} - \frac{1}{4}  T_{\mu \alpha} 
T^{\alpha}_{\nu} + \frac{1}{24} g_{\mu \nu} \left[ 3 T_{\alpha \beta} 
T_{\alpha \beta} - (T^{\alpha}_{\alpha})^{2} \right] .
\end{equation}
where $T = T^{\alpha}_{\alpha}$, $S_{\mu \nu}$ is significant at high 
energies i.e., $\rho > \lambda$, 

$\bullet$
$E_{\mu \nu} $ :  occurs  due to the non-local effects from the 
free gravitational field in the bulk, which  enters  in the equation via the
 projection 
${\bf E}^{(5)}_{AB} =  C^{(5)}_{ACBD} n^{C} n^{D}$ where $n^{A}$ is 
normal to the surface ($n^{A} n_{A} = 1$). The term is symmetric and 
traceless 
and without components orthogonal to the brane, so ${\bf E}_{AB} n^{B} = 0$
and ${\bf E}_{AB} \rightarrow  E_{\mu \nu} g^{\mu}_{A} g^{\nu}_{B} $
as $y \rightarrow 0$

The  Robertson-Walker metric in the brane world is
\begin{equation}
ds^{2} = - dt^{2} + a^{2}(t) \left[ dr^{2} + \Sigma_{k}^{2}(r) ( d\theta^{2} + sin^{2} \theta \; \phi^{2} ) \right] 
\end{equation}
where $ \Sigma_{k}^{2}(r) =   sin \, r (sinh \, r) $ for $k = 1 (-1)$ and $\Sigma_{k} = r $ for $ k = 0$ and $a(t)$ is the scale factor. In this paper we consider brane world described by a flat Robertson-Walker metric. We also set cosmological constant to be zero.
The matter in it is assumed to be 
equivalent to that of a perfect fluid and therefore the energy momentum tensor is
\begin{equation}
T_{\mu \nu} = (\rho +  p ) u_{\mu} u_{\nu} + p g_{\mu \nu} 
\end{equation}
where $u^{\mu} u_{\mu} = - 1$, $u^{\mu}$ is the unit fluid velocity of matter and $\rho$ is the energy density and $p$ is the pressure of the matter fluid.

Using the metric (6) and  the energy momentum tensor (7), the field equation (2) for a flat brane metric become
\begin{equation}
H^{2} = \frac{\kappa^{2}}{ 3 } \; \rho  \left( 1 + \frac{\rho}{2 \lambda} \right)  + \frac{\it E}{a^{4}}
\end{equation}
where $\kappa^2 = 8 \pi G = 8 \pi/M_{4}^{2}$ and $\it E$ is the  black radiation term. The second term in the above equation is the quadratic correction in energy density $\rho$. In the high energy limit, $\lambda \rightarrow \infty$ and $E = 0$ i.e., neglecting dark energy one recovers the standard four dimensional general relativistic results. In the very early universe the quadratic term in energy density dominates over the other terms giving rise to a new kind of variation of the Hubble parameter with  the energy density on the brane. The conservation equation is
\begin{equation}
\frac{d\rho}{dt} + 3 H (\rho + p) = 0 .
\end{equation}
We now consider scalar field minimally coupled to gravity to describe the matter content in the universe. For a homogeneous scalar field $\phi = \phi (t)$ the energy density and pressure are given by 
\[
\rho = \frac{1}{2} \dot{\phi}^{2} + V(\phi),
\]
\begin{equation}
p = \frac{1}{2} \dot{\phi}^{2} - V(\phi).
\end{equation}
In the four dimensional general theory of relativity the condition for inflation is   $\frac{1}{2} \dot{\phi}^{2} << V(\phi)$  i.e., $p < - \frac{1}{3} \rho$. Maartens {\it et al.} [7] obtained chaotic inflation scenario on the brane with this condition using a massive scalar field. It is known [8, 9] that chaotic scenario can be realized even if  $\frac{1}{2} \dot{\phi}^{2} >>  V(\phi)$ in GTR. In this paper we consider two possibilities : (i)   $\frac{1}{2} \dot{\phi}^{2} >> V(\phi)$   and (ii) $\frac{1}{2} \dot{\phi}^{2} << V(\phi)$. In the first case we also look for subsequent inflation.

{\bf III. COSMOLOGICAL SOLUTIONS }

{\bf CASE I :} Let us consider first the case  $\frac{1}{2} \dot{\phi}^{2} >> V(\phi)$, which is reasonable at early universe. In this case one obtains $\rho = p =  \frac{1}{2} \dot{\phi}^{2}$ which corresponds to stiff matter. Equation (9) therefore can be written as 
\begin{equation}
\ddot{\phi} + 3 H \dot{\phi}  = 0.
\end{equation}
On integrating we get 
\begin{equation}
\dot{\phi} =  \pm \frac{C}{a^{3}}
\end{equation}
where $C$ is an integration constant. Using the above relation we write
\begin{equation}
\left( \frac{\dot{a}}{a} \right)^{2} = \frac{\alpha^{2}}{a^{6}} \left(1 + \frac{\beta^{2}}{a^{6}} \right) 
\end{equation}
where we use the symbols for simplification $\alpha = \sqrt{\frac{4 \pi C^{2}}{3M_{4}^{2}}}$ and   $\beta^{2} = \frac{C^{2}}{4 \lambda}$.
Equation (13) can be easily integrated to obtain
\begin{equation}
a(t) =  \left[ \left( 3 \alpha t + A \right)^{2} - \beta^{2} \right]^{1/6}  \end{equation}
where $A$ is an integration constant. Two models are possible depending upon the boundary conditions : (i) universe  from a singularity when $A = \beta$ and (ii) universe without singularity when $A = \beta^{2} + a_{o}^{6}$. We discuss below two models of the universe.

{\it Model I :  Universe from singularity }

In this case we set $A = \beta$. The evolution of the the scale factor of the universe is given by 
\begin{equation}
a(t) =  \left(9 \alpha^{2} t^{2} + 6 \alpha \beta t \right)^{1/6} . \end{equation}
In the high energy limit, $\beta = \frac{C^{2}}{4 \lambda} \rightarrow \infty$ i.e., the second term inside the bracket is comparatively larger one,  therefore the scale factor evolves as
\begin{equation}
a(t) =  \left( 6 \alpha \beta \right)^{1/6} t^{1/6},  
\end{equation}
and  consequently the scalar field evolves as 
\begin{equation}
\phi(t) =  \phi_{o} - \frac{ 2 C}{\sqrt{6 \alpha  \beta}} t^{1/2}.
\end{equation}
However
in the low energy limit $\beta = \frac{C^{2}}{4 \lambda} \rightarrow 0$, one gets  the scale factor which evolves as
\begin{equation}
a(t) =  \left( 3 \alpha \right)^{1/3} t^{1/3},  
\end{equation}
and the corresponding scalar field evolution is given by 
\begin{equation}
\phi(t) =  \phi_{o} - \frac{ 2 C}{\sqrt{6 \alpha}} \; \ln t.
\end{equation}
The  solutions obtained  above for two different energy limits have different rates of expansion at the early epoch. The physical implications of this behaviour may be realised  from the point of view of singularity, in the braneworld it  approaches more  slowly compared to that in  GTR which is shown below
\[
\frac{1}{a} \left( t \frac{da}{dt} \right)|_{Brane} = \frac{1}{6} < \frac{1}{a} \left( t \frac{da}{dt} \right)|_{GTR} = \frac{1}{3}.
\]
 However, one finds that the pre-inflationary stage continues for time 
 $t <<  t_{i} = \left( \frac{3 \alpha}{16 \lambda b_{o}} \right)^{1/3}$ in brane with a self interacting scalar field described by a potential $V(\phi) = \frac{1}{4} b_{o} \phi^{4}$, where $b_{o}$ represents the coupling constant.  It is evident that the kinetic energy decreases more rapidly than the potential energy and at $t = t_{i}$ the potential energy becomes important which leads to inflation. The inflationary scenario may be realized with a polynomial potential $V(\phi) \sim \phi^{n}$, with $n > 1$.

{\it Model II :  Universe without singularity }

Here we consider power law behavior of the scale factor  given by (14), with $a(t=0) = a_{o} = (A^{2} - \beta^{2})^{1/6}$ with $A \neq \beta $.   One gets a singularity free universe with a scalar field which evolves  as $\phi = \phi_{o}  \pm \sqrt{\frac{2C^2}{3 \alpha \beta}} \; t$. However in the low energy limit, the scalar field evolves more slowly $\phi = \phi_{o}  \pm \frac{C}{3 \alpha} \; ln (3 \alpha t + A) $. 

The pre-inflationary stage in the above two models of the brane-world ends very soon as the kinetic energy density decreases faster than the potential energy and the universe transits to an inflationary stage which is discussed in the next section. 

{\bf CASE II :} The case $\frac{1}{2} \dot{\phi}^{2} << V(\phi)$ is studied by Maartens {\it et al.} [7] for a massive scalar field. In this paper  we consider a self interacting scalar field with  a potential $  V(\phi) =  
\frac{1}{4} b_{o} \phi^{4} $, where $b_{o}$ is the coupling strength of the scalar field. The field equation can be written as 
\begin{equation}
\left( \frac{\dot{a}}{a} \right)^{2} = \frac{2 \pi b_{o} }{3 M_{4}^{2}}  \phi^{4} \left( 1 + \frac{b_{o}}{8 \lambda} \phi^{4} \right)
\end{equation}
and the scalar field equation becomes
\begin{equation}
3 H \dot{\phi} = - b_{o} \; \phi^{3}. 
\end{equation}
Equations (20) and (21) are used to determine the scale factor of the universe which is given below
\begin{equation}
a(t) =  a_{o} e^{\left[ \frac{\pi}{M_{4}^{2}} \left( \phi_{o}^{2} - \phi^{2} \right)  + \frac{\pi b_{o}}{24 \lambda M_{4}^{2}} \left(\phi_{o}^{6} - \phi^{6} \right) \right]}.
\end{equation}
In the low energy limit
$\frac{ b_{o}}{\lambda} \rightarrow 0$, one recovers the solution obtained by Linde [8]. However, in the high energy limit $\frac{ b_{o}}{\lambda} \rightarrow  \infty$, we get brane cosmological solution where the scale factor of the universe grows as  
\begin{equation}
a(t) =  a_{o} e^{\frac{\pi b_{o}}{24 \lambda M_{4}^{2}} \left(\phi_{o}^{6} - \phi^{6} \right)}.
\end{equation}
The scalar field evolves as 
\begin{equation}
\phi = \left[ \phi_{o}^{2} -  \sqrt{\frac{ 16 \lambda M_{4}^{2}}{3 \pi}} \; t \right]^{1/2}.
\end{equation}
It is evident from  the scale factor obtained in (23)  that a huge inflation results in the brane as  $\frac{ b_{o}}{\lambda}$ is very large. It is important to determine the initial value of the scalar field to get sufficient inflation in order to  explain the homogeneity and isotropy of the observed universe. 
At the end of inflation $\phi_{f} = \phi_{end}$, one obtains the number of e-folds of inflationary expansion, $N_{COBE}  = \int_{t_{i}}^{t_{f}} H  dt \sim 65 $. To get sufficient inflation in this case we  obtain   $\phi_{COBE} \approx \left( \frac{1170}{ \pi^{2} b_{o}} \right)^{1/6} M_{5}$.  The amplitude of scalar perturbation during slow-roll inflation may be evaluated which is given by
\begin{equation}
A_{S}^{2} \sim \left( \frac{512 \pi}{75 M_{4}^{6}}  \right)\frac{V^{3}}{V'^{2}} \left[
\frac{2 \lambda + V}{ 2 \lambda} \right]^{3}|_{k = aH}  \sim \frac{\pi^{4}  b_{o}^{4}}{2025} \frac{\phi^{18}}{M_{5}^{18}}. 
\end{equation}
Using the result of COBE normalization [15] $A_{S} \sim 2 \times 10^{- 5}$, we estimate the coupling constant of the scalar field which is $b_{o} \sim 5.06 \times 10^{-16}$. Thus  $\phi_{COBE}$ fixes  the self interacting coupling constant of the inflaton field. It may be pointed out here that this is a tighter constraint compared to the usual 4D one. Thus in the brane world fine tunning problem is not very much improved compared to that   in 4D FRW cosmology.
The spectral index of the scalar spectrum ${\it n}_{s} - 1 = \frac{d \ln A_{S}^{2}}{d \ln k}|_{k = a H} $, is given by  \begin{equation}
{\it n}_{s} - 1 = - \frac{576}{k} \left( \frac{\lambda}{b_{o}} \right) \phi_{COBE}^{12} .
\end{equation}
In the high energy limit $\frac{b_{o}}{\lambda} \rightarrow \infty$, which we are discussing here leads to spectral index very close to unity ( ${\it n}_{s} \rightarrow 1$ ).
 In GTR, Linde [8] suggested that the observed part of the universe originates from a fluctuation region with $\phi (t = M_{4}^{-1}) > 3 M_{4}$ where $M_{4} \sim 1.2 \times 10^{19} GeV$. The lower bound of $\phi$ ensures that there is   sufficient inflation, which however becomes $\phi > 7.85 \times 10^2 M_{5}$ in the brane world scenario. As $M_{5} < 10^{17} $ GeV, chaotic inflationary universe may be obtained from an initial distribution of scalar field having values even below the Planck mass scale.

{\bf IV. Discussions :}

We studied  chaotic inflationary universe model in brane world considering a minimally coupled self-interacting  homogeneous scalar field. It is found that the chaotic inflation may be  realized when  the potential energy of the inflaton field dominates  the kinetic energy to begin with. It is also found that when kinetic energy dominates initially, a pre-inflationary phase results, which subsequently transits to inflationary universe as the kinetic energy decreases  much faster than the potential energy.   The universe evolves out of a chaotic distribution of initial data which may  not necessarily have values greater than Planck mass scale.  In the above a sufficient inflation is obtained with a scalar field $\phi_{i} > 7.85 \times 10^{2} M_{5}$ which is less than that required in the 4D case.
A very weakly coupled ($b_{o} \sim 10^{- 16}$) scalar field is required on the brane compared to that ($b_{o} \sim 10^{- 12}$) required in GTR which, however, worsen the fine tunning problem. We also note that in the  case of a universe created with a singularity, the rate of approach of singularity on the brane is much faster than that one gets in  GTR. 
\vspace{0.5in}

{\large \it Acknowledgement :}
I would like to  thank the Inter-University Centre
for Astronomy and Astrophysics (IUCAA) Pune for providing a facility where this work was carried out. I would also like to thank Prof. S. Mukherjee for providing facilities to work at IUCAA Reference Centre at  North Bengal University.
\pagebreak

\end{document}